\documentclass[12pt]{article}
\usepackage{graphicx}
\usepackage{subfig}
\usepackage{placeins}
\usepackage{hyperref}

\newcommand\pubnumber{ATL-PHYS-PROC-2022-114}
\newcommand\pubdate{\today}

\newcommand\mtop{\ensuremath{m_t}}
\newcommand\invfb{\ensuremath{\mathrm{fb^{-1}}}}
\newcommand\ttbar{\ensuremath{t\bar{t}}}
\newcommand\pt{\ensuremath{p_{\mathrm{T}}}}
\newcommand\transmass{\ensuremath{m_{\mathrm{T}}}}

\makeatletter
\def\blfootnote{\xdef\@thefnmark{}\@footnotetext}
\makeatother

\def\institute{INFN Sezione di Roma Tor Vergata}
\def\support{\blfootnote{Copyright 2022 CERN for the benefit of the ATLAS and CMS collaborations.\\ CC-BY-4.0 license.}}

\def\Title#1{\begin{center} {\Large #1 } \end{center}}
\def\Author#1{\begin{center}{ \sc #1} \end{center}}
\def\Address#1{\begin{center}{ \it #1} \end{center}}

\newcommand\pubblock{\rightline{\begin{tabular}{l} \pubnumber\\
         \pubdate  \end{tabular}}}
\newenvironment{Abstract}{\begin{quotation}  }{\end{quotation}}
\newenvironment{Presented}{\begin{quotation} \begin{center} 
             PRESENTED AT\end{center}\bigskip 
      \begin{center}\begin{large}}{\end{large}\end{center} \end{quotation}}

\def\beq{\begin{equation}}
\def\eeq#1{\label{#1}\end{equation}}
\def\eeqn{\end{equation}}

\def\beqa{\begin{eqnarray}}
\def\eeqa#1{\label{#1}\end{eqnarray}}
\def\eeqan{\end{eqnarray}}

\let\bar=\overbar

\def\Dslash{\not{\hbox{\kern-4pt $D$}}}
\def\dslash{\not{\hbox{\kern-2pt $\del$}}}

\def\msb{{\bar{\ssstyle M \kern -1pt S}}}

\begin{document}
\begin{titlepage}
\pubblock

\vfill
\Title{Direct top quark mass measurements with the ATLAS and CMS detectors}
\vfill
\Author{ Marco Vanadia, on behalf of the ATLAS and CMS collaborations\support}
\Address{\institute}
\vfill
\begin{Abstract}
This article presents a review of recent results on direct top quark mass measurements performed by the ATLAS and CMS collaborations on $pp$ collisions collected during Run 2 of the LHC at $\sqrt{s}=$13~TeV.
\end{Abstract}
\vfill
\begin{Presented}
$15^\mathrm{th}$ International Workshop on Top Quark Physics\\
Durham, UK, 4--9 September, 2022
\end{Presented}
\vfill
\end{titlepage}
\def\thefootnote{\fnsymbol{footnote}}
\setcounter{footnote}{0}

\section{Introduction}

The top quark is the heaviest elementary particle in the Standard Model (SM) of particle physics. A precise measurement of its mass \mtop{} is crucial for Electroweak precision tests and for predictions for physics phenomena within the SM and beyond.
A review of recent direct measurements of \mtop{} at the LHC~\cite{LHC} is presented here. These measurements reconstruct \mtop{} from the kinematics of the decay products of the top quark. 
Typically they fit an observable measured in data with templates generated as a function of the top quark mass parameter of the Monte Carlo (MC) generators.
The ATLAS~\cite{ATLAS} and CMS~\cite{CMS} collaborations published the most precise direct \mtop{} measurements of 172.69$\pm$0.48 and 172.44$\pm$0.49~GeV~\cite{ATLASR1,CMSR1}, obtained by combining several measurements in different signatures performed on $pp$ collisions collected during Run 1 of the LHC. 
A combined value of 174.30$\pm$0.65~GeV has been obtained by Tevatron experiments~\cite{TEVATRON}. 
The results reviewed in this article are obtained on $\sqrt{s}=$13~TeV $pp$ collisions from Run 2 of the LHC.

\section{Measurements}

\begin{figure}[htb]
\centering
\subfloat[ ]{\includegraphics[width=0.8\linewidth]{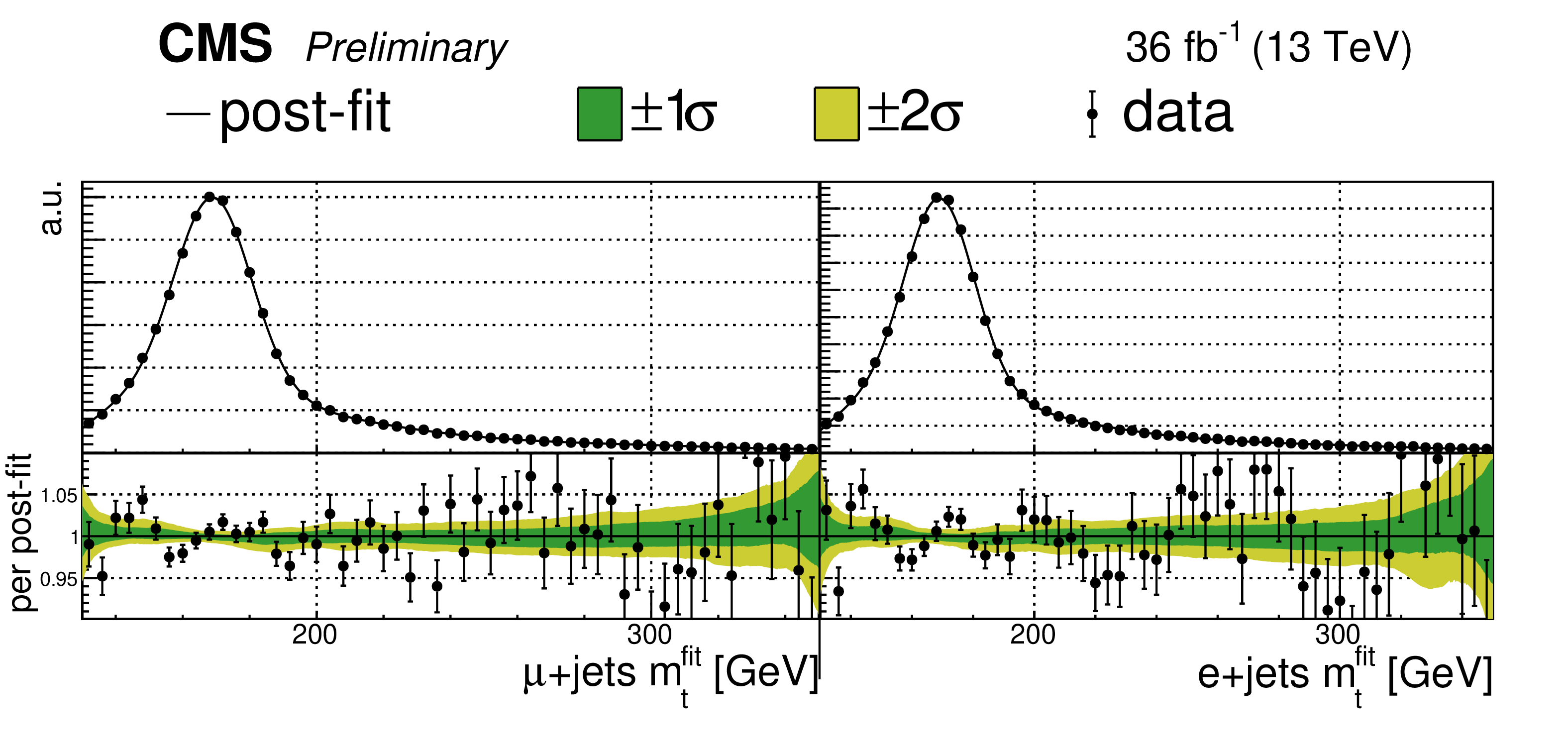}}
\caption{Post-fit distribution of $\mtop^{\mathrm{fit}}$ in the (left) muon and (right) electron channels for the measurement in Ref.~\cite{CMSljets}.}
\label{fig:CMSljets}
\end{figure}

A \mtop{} measurement by the CMS Collaboration on a data set of 36~\invfb{} in \ttbar{} $\ell$+jets events is reported in Ref.~\cite{CMSljets}. Events with exactly 1 isolated electron or muon
and at least 4 hadronic jets 
are selected.  
Exactly 2 jets are required to be identified as $b-$jets by a tagging algorithm. A $\chi^2$ quantity is defined using the measured kinematic properties of the reconstructed objects and their resolution, 
imposing the $W$ boson mass $m_{W}$ as a constraint; \mtop{} is a free parameter of the fit.
The fitted top mass $\mtop^{\mathrm{fit}}$ for events with goodness-of-fit GOF$>$0.2 is the main observable used in the analysis. The measurement is improved by using additional observables for those events: the reconstructed $m_{W}^{\mathrm{reco}}$, the ratio between the invariant mass of the $\ell+b-$jet system $m_{\ell b}$ and $\mtop^{\mathrm{fit}}$, 
and the ratio $R^{\mathrm{reco}}_{bq}$ between the scalar sum of the \pt{} of the $b-$jets and the scalar sum of the \pt{} of the light-jets.
These provide sensitivity to the jet energy scale. A fifth observable used in the analysis is $m_{\ell b}$ for events with GOF$<$0.2. Backgrounds are obtained from simulations.
The \mtop{} is measured with a Maximum Likelihood (ML) profiled fit. Systematic uncertainties are treated as nuisance parameters (NPs); with this technique the data is used to obtain more information on the uncertainties and possibly to constrain them, even if additional studies are typically needed to support eventual strong constraints.
The result of the measurement, shown in Fig.~\ref{fig:CMSljets}, is \mtop$=171.77\pm0.38$~GeV, the measurement with the smallest uncertainty to date. The uncertainty is dominated by systematic effects; the most relevant ones are associated with the $b-$jet energy measurement and the modelling of the final state radiation (FSR) and color reconnection (CR) in simulation. The measurement is lower than the world average based on previous measurements, but this is mostly associated with the correlation model chosen for the FSR uncertainty, where variations for different branching types ($g\rightarrow gg$, $g\rightarrow qg$, $q\rightarrow qg$ for $q=u,d,s,c$ and $q\rightarrow qg$ for $q=b,t$) are kept uncorrelated. It is shown that by correlating those variations, as done in previous measurements, the difference with those results can be significantly reduced.\\

\begin{figure}[htb]
\centering
\subfloat[ ]{\includegraphics[width=0.49\linewidth]{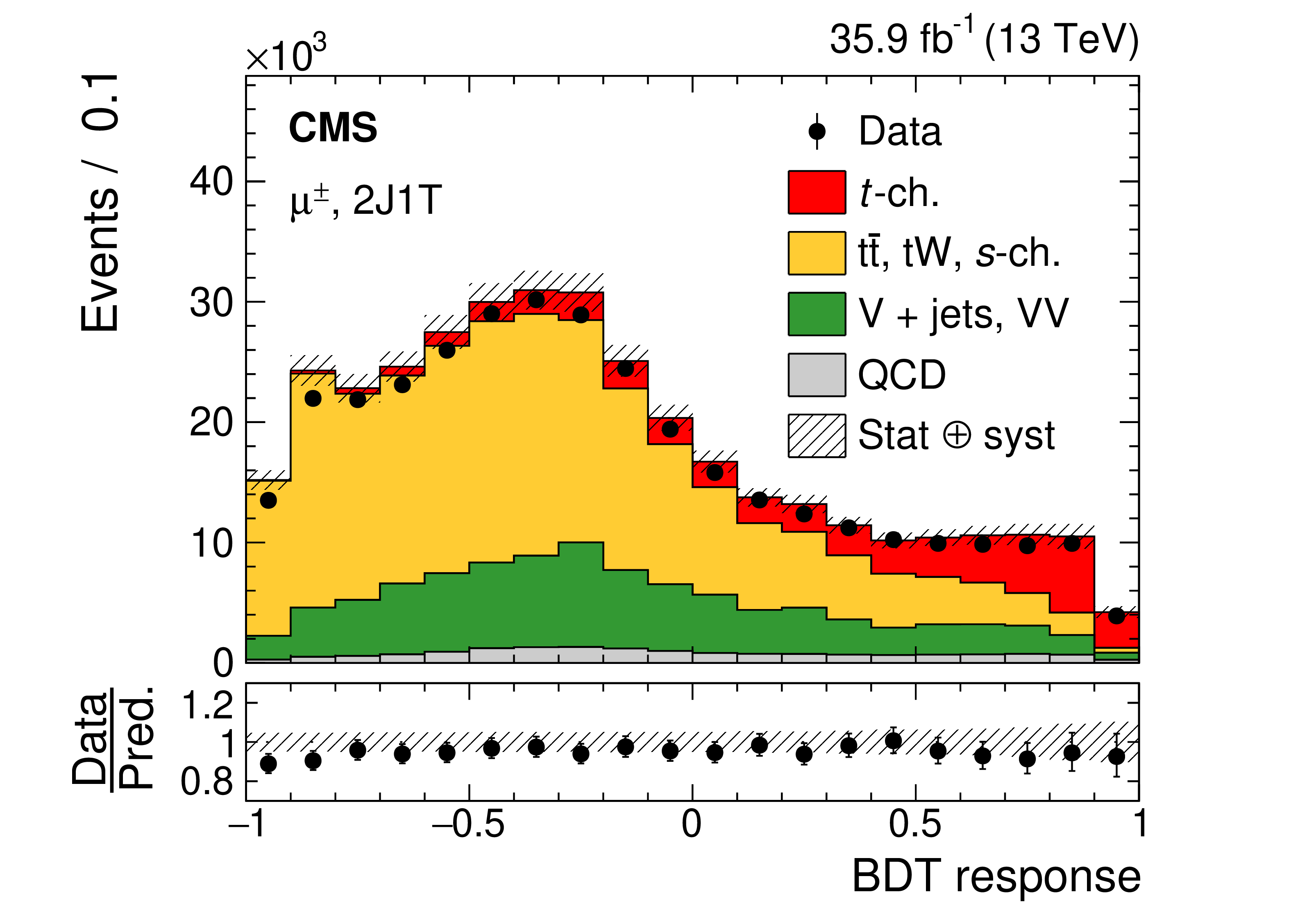}}
\subfloat[ ]{\includegraphics[width=0.49\linewidth]{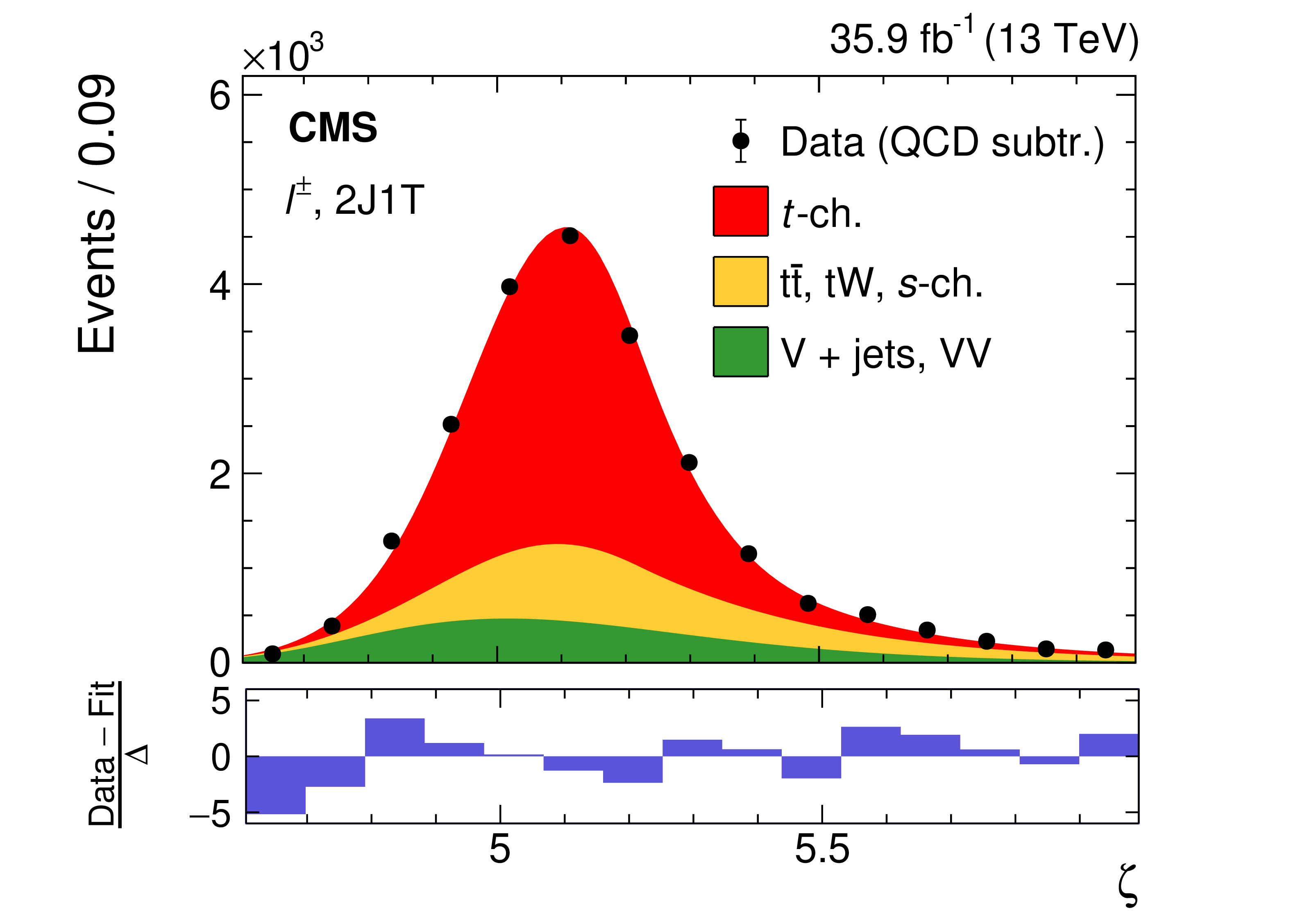}}
\caption{(a) Distribution of the BDT score in the muon channel and (b) fit to the final observable for the measurement in Ref.~\cite{CMSstop}}
\label{fig:CMSstop}
\end{figure}

A measurement performed by the CMS collaboration on a data set of 36~\invfb{} in the single-top $t-$channel is published in Ref.~\cite{CMSstop}. This channel has lower statistics and purity than \ttbar{} ones, but it explores a different kinematic region and is subject to partially independent systematic uncertainties. Events with exactly 1 isolated electron or muon and exactly 2 jets are selected. Exactly 1 jet is required to be identified as a $b-$jet by a dedicated algorithm; events with no $b-$jets are used to define a validation region for the analysis. The transverse mass of the charged lepton plus neutrino system \transmass{} is required to be greater than 50 GeV, to reduce background contributions. The QCD multijet background is controlled using a data-driven technique, exploiting events with non-isolated leptons and events with low \transmass{}. Other backgrounds are obtained from simulations. \mtop{} is directly reconstructed using the kinematic properties of the selected objects, with the momentum of the neutrino inferred from $\vec{p}_{\mathrm{T}}^{~\mathrm{miss}}$ and imposing $m_{W}$ as a constraint.
In order to improve the signal-over-background ratio, a Boosted Decision Tree (BDT) method is defined, using several kinematic quantities as input, of which the most powerful one is the angular separation between the two jets in the event. For the final selection events are required to be associated with a BDT score greater than 0.8, a value optimised to reduce the total uncertainty of the measurement. The final observable is defined as $\zeta=ln(\mtop^{\mathrm{reco}}/\mathrm{1~GeV})$.
After the QCD multijet background is subtracted, the other contributions, modelled with analytical functions, are fitted to the data, as shown in Fig.~\ref{fig:CMSstop}. The peak position of the distribution is sensitive to the \mtop{} value, which is measured to be $\mtop=172.13^{+0.76}_{-0.77}$~GeV. The uncertainty is dominated by systematic effects associated with the jet energy scale (JES), $b-$jet identification, and modelling of FSR, CR and $b-$hadronization effects.\\

\begin{figure}[htb]
\centering
\subfloat[ ]{\includegraphics[width=0.4\linewidth]{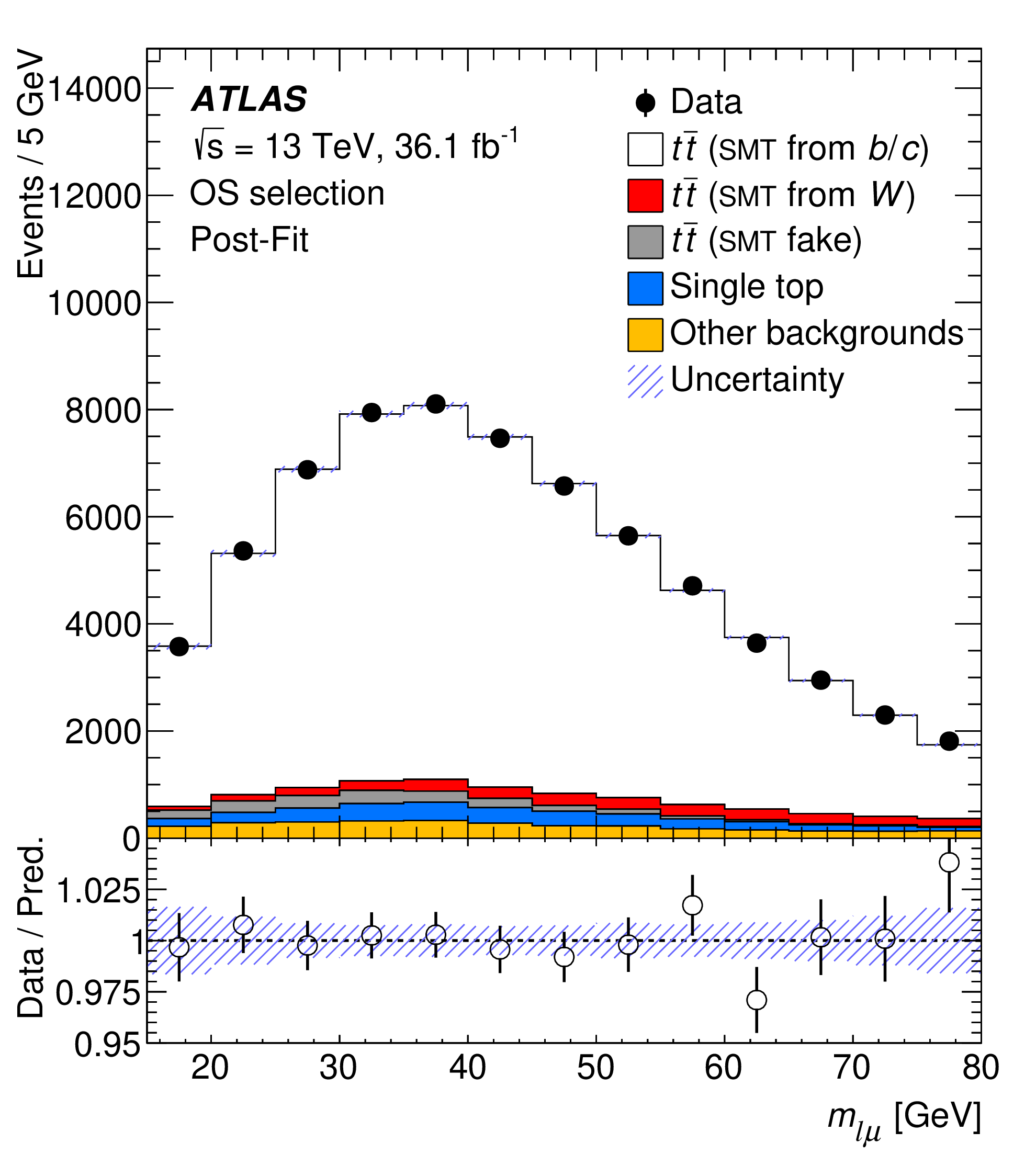}}
\subfloat[ ]{\includegraphics[width=0.4\linewidth]{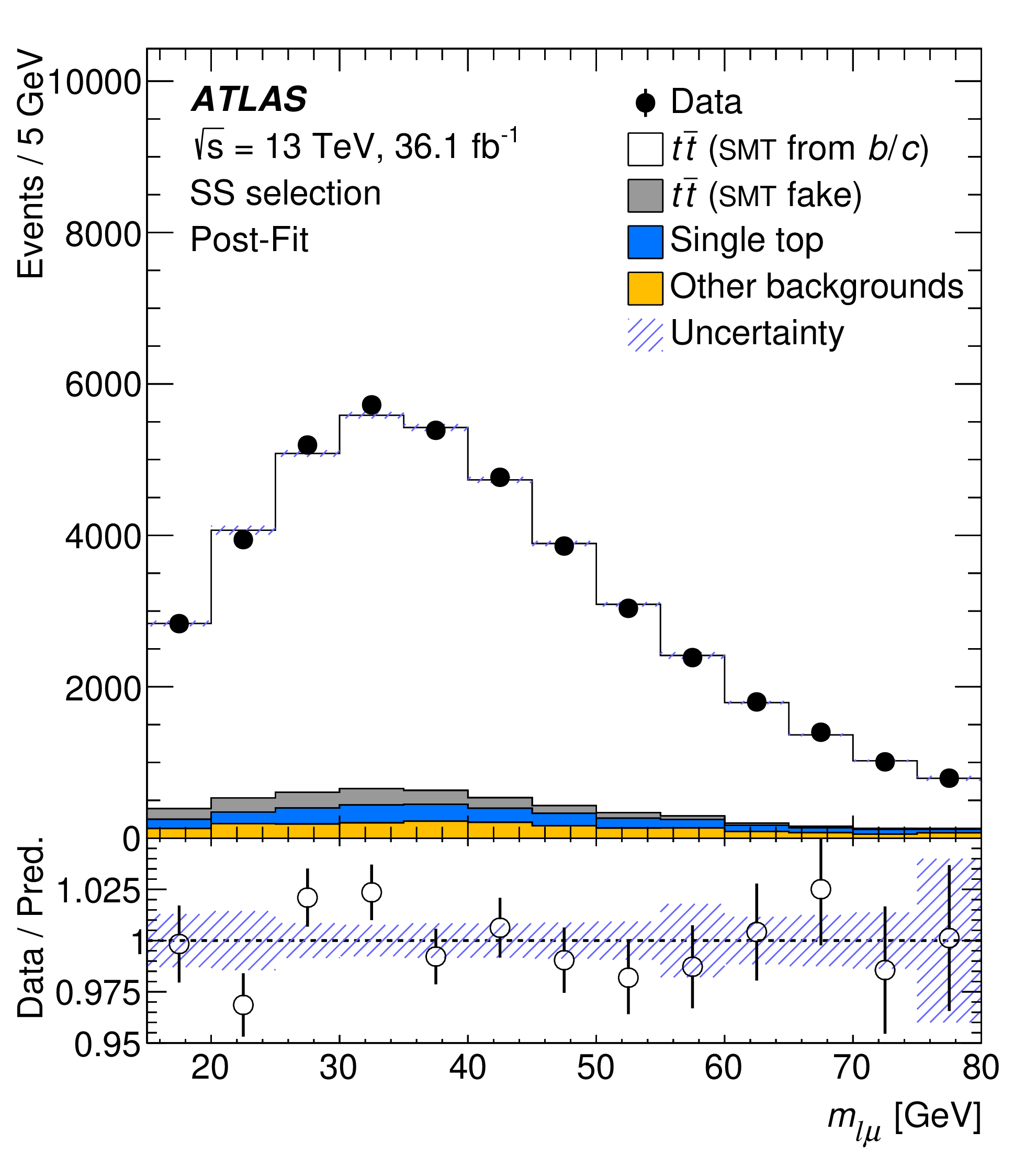}}
\caption{Post-fit distribution of $m_{\ell\mu}$ for events with an isolated lepton and a SMT muon with charges with (a) opposite sign (OS) and (b) same sign (SS) for the measurement in Ref.~\cite{ATLASsmt}}
\label{fig:ATLASsmt}
\end{figure}

Ref.~\cite{ATLASsmt} presents a measurement performed by the ATLAS collaboration on a data set of 36~\invfb{} in \ttbar{} $\ell$+jets events with a $b-$hadron decaying semileptonically to a muon, $B \rightarrow \mu +X$. The invariant mass of the isolated lepton and of the muon produced by the $b-$hadron decay, $m_{\ell\mu}$, is a purely leptonic observable sensitive to the \mtop{} value. It is largely insensitive to jet-related uncertainties, providing a complementary sensitivity with respect to jet-based measurements. Events with exactly 1 isolated $e$ or $\mu$ and at least 4 jets are selected, one of which must be identified as a $b-$jet by a dedicated algorithm. $B \rightarrow \mu +X$ decays are identified using a soft muon tagging (SMT) algorithm, requiring that a muon (SMT muon) is identified within $\Delta R<0.4$ from the axis of one the selected jets. $V+$jets and QCD multijet backgrounds are determined using data-driven techniques, while other backgrounds are modelled using MC samples. A detailed study is performed on $e^{+}e^{-}\rightarrow Z\rightarrow b\bar{b}$ data to tune the parameters controlling the $b-$quark fragmentation modelling in MC. This procedure assumes the universality of the $b-$fragmentation for lepton and hadron colliders, as supported by existing data in the phase space relevant for the analysis, within the quoted uncertainties. A ML profiled fit is performed on $m_{\ell\mu}$ using systematic uncertainties as NPs, as shown in Fig.~\ref{fig:ATLASsmt}. The final result is \mtop$=$174.41$\pm$0.39~(stat.)$\pm$0.66~(syst.)$\pm$0.25~(recoil)~GeV. Dominant systematic uncertainties are associated with the modelling of the $b-$hadron decay. The recoil uncertainty is associated with an intrinsic ambiguity in the definition of the recoil for gluon radiation from the $b-$quark in $t\rightarrow Wb$ decays for the signal MC simulation. This uncertainty was developed after input from the MC authors and will need to be carefully addressed by future measurements.\\

\begin{figure}[htb]
\centering
\subfloat[ ]{\includegraphics[width=0.45\linewidth]{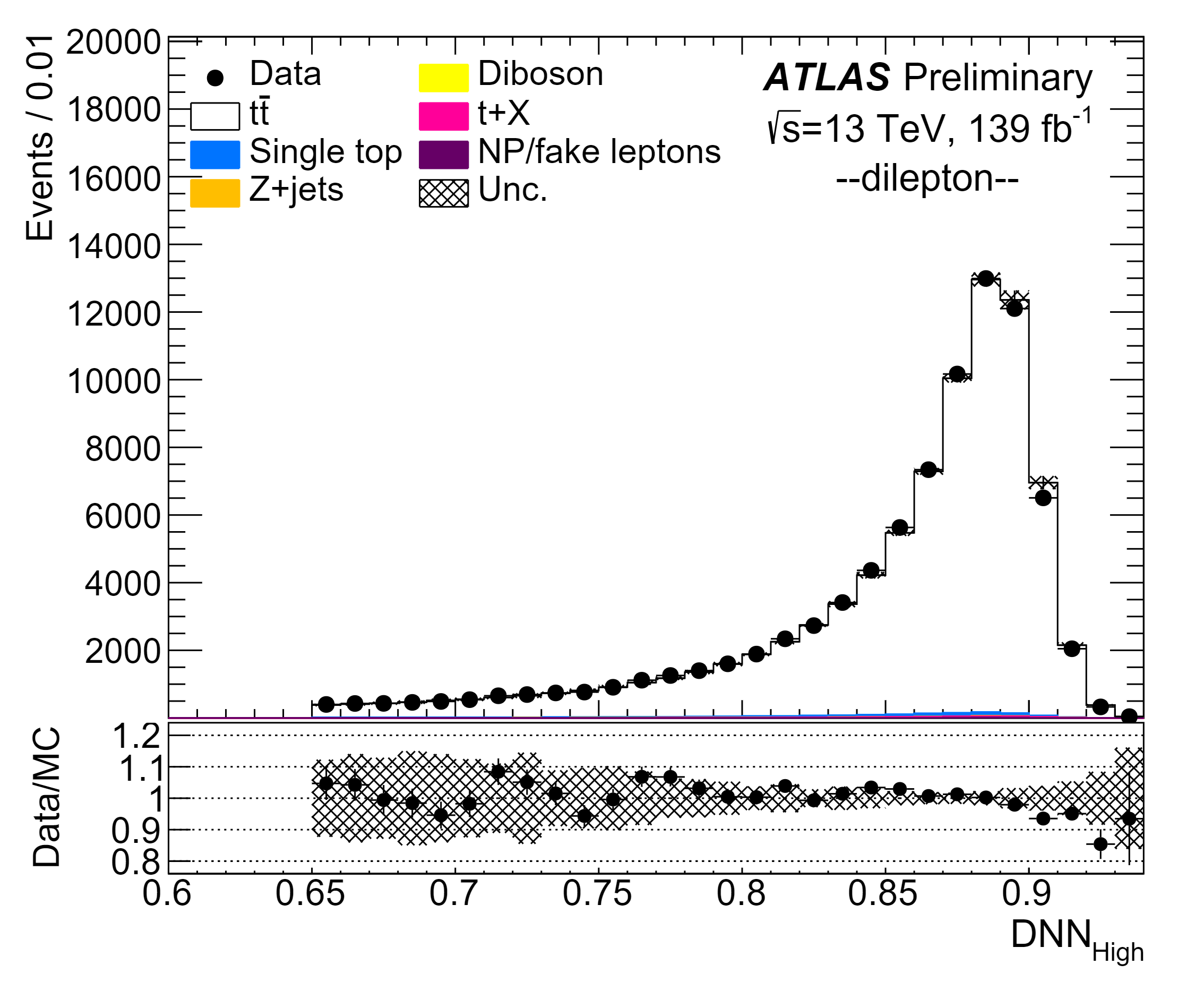}}
\hspace{1cm}
\subfloat[ ]{\includegraphics[width=0.4\linewidth]{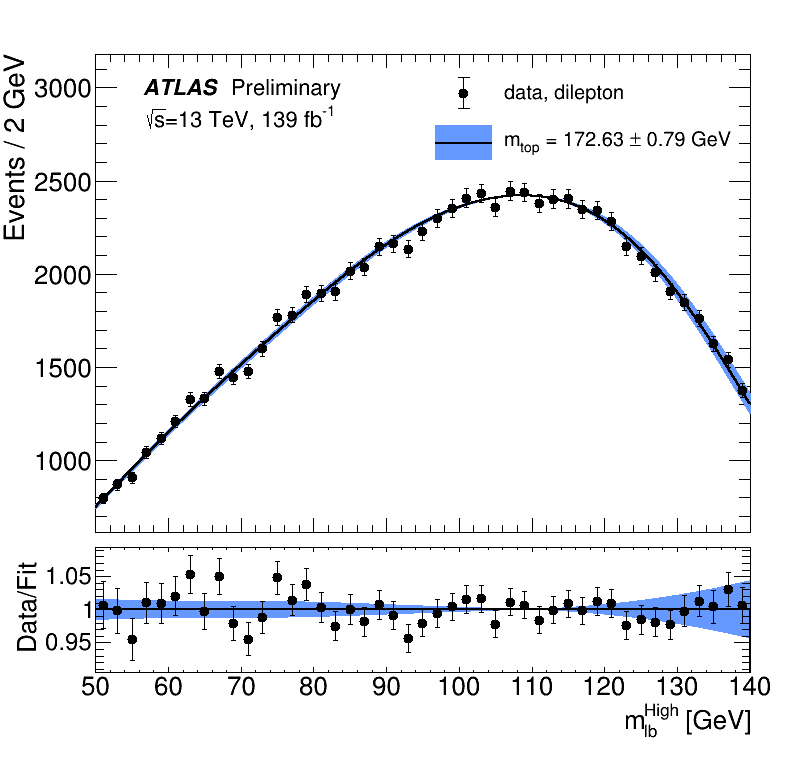}}
\caption{(a) DNN score for the selected $b-$jet/$\ell$ pair  and (b) $m_{\ell b}^{High}$  distribution for the measurement in Ref.~\cite{ATLASdil}}
\label{fig:ATLASdil}
\end{figure}

Finally, Ref.~\cite{ATLASdil} presents a measurement by the ATLAS collaboration on the full Run 2 data set (139~\invfb{}) in the \ttbar{} $\ell\ell$ channel. Events with two oppositely charged $e/\mu$, and at least two jets, are selected. Exactly two jets are required to be identified as $b-$jets by a tagging algorithm. A veto is applied in the $ee$ and $\mu\mu$ channels on the invariant mass of the two leptons $m_{\ell\ell}$, to suppress contributions from decays of known resonances. A new method, based on a Deep Neural Network (DNN), is used to reconstruct the event topology by finding the appropriate $b-$jet/$\ell$ pairing, using several kinematic quantities as an input. Events with a high probability of being correctly reconstructed by the DNN are selected. Then the $b-$jet/$\ell$ pair with the highest \pt{} is used in the rest of the analysis, and only events where 
the $b-$jet in the pair is the leading one in the event are retained. A final selection requiring the \pt{} of the pair to be greater than 160~GeV is applied. The modelling of this variable yields a significant difference between MC and data, which is at least partly expected due to missing higher order corrections to the top quark \pt{}; a careful study showed that the impact of such mismodelling is well controlled within the uncertainties quoted for the measurement. The invariant mass $m_{\ell b}^{\mathrm{High}}$ of that pair is used as the final observable, and an unbinned ML fit is performed to extract the \mtop{} as shown in Fig.~\ref{fig:ATLASdil}. This results in \mtop$=$172.63$\pm$0.20~(stat.)$\pm$0.67~(syst.)$\pm$0.37~(recoil)~GeV, with dominant uncertainties associated with modelling effects (matrix element generator, ISR/FSR, CR) and JES. The recoil uncertainty, discussed above, is taken into account for the first time for a jet-based measurement; without it, the result improves by 17\% with respect to the precision of the one based on Run 1 data. An additional check performed on non-resonant and off-shell effects showed that these are well within the quoted modelling uncertainty.\\

\section{Conclusion}

A review of recent direct top quark measurements by the ATLAS and CMS collaborations on LHC Run 2 data was presented. The measurements reach a level of precision comparable with that of the theoretical uncertainties on the definition of \mtop{} itself, posing a significant challenge on their interpretation~\cite{REVIEW}. Furthermore, they are now limited by systematic uncertainties associated with the modelling in MC simulations, which inevitably introduces a dependency of the experimental results on the specific simulation. With the LHC providing already more data with the ongoing Run 3, and the potential of the Run 2 data set not yet fully explored, the next years will pose a great challenge for experimental and theoretical physicists, which are called on to build a solid framework for improving and interpreting the \mtop{} measurements.

\FloatBarrier


\begin{thebibliography}{99}

\bibitem{LHC}
L. Evans and P. Bryant (editors), LHC Machine, 2008 JINST 3 S08001

\bibitem{ATLAS}
ATLAS Collaboration, The ATLAS Experiment at the CERN Large Hadron Collider, 2008 JINST 3 S08003

\bibitem{CMS}
CMS Collaboration, The CMS experiment at the CERN LHC, 2008 JINST 3 S08004

\bibitem{ATLASR1}
ATLAS Collaboration, Measurement of the top quark mass in the $\ttbar \rightarrow$lepton+jets channel from $\sqrt{s}=$8 TeV ATLAS data and combination with previous results, EPJC 79 (2019) 290

\bibitem{CMSR1}
CMS Collaboration, Measurement of the top quark mass using proton-proton data at $\sqrt{s}=$7 and 8 TeV, PRD 93 (2016) 072004

\bibitem{TEVATRON}
The Tevatron Electroweak Working Group, Combination of CDF and D0 results on the mass of the top quark using up to 9.7 \invfb{} at the Tevatron, (2016), arXiv:1608.01881

\bibitem{CMSljets}
CMS Collaboration, A profile likelihood approach to measure the top quark mass in the lepton+jets channel at $\sqrt{s}=$13 TeV, CMS-PAS-TOP-20-008, \url{https://cds.cern.ch/record/2806509}

\bibitem{CMSstop}
CMS Collaboration, Measurement of the top quark mass using events with a single reconstructed top quark in $pp$ collisions at $\sqrt{s}=$13 TeV, JHEP 12 (2021) 161

\bibitem{ATLASsmt}
ATLAS Collaboration, Measurement of the top-quark mass using a leptonic invariant mass in $pp$ collisions at $\sqrt{s}=$13 TeV with the ATLAS detector, arXiv:2209.00583 (submitted to JHEP)

\bibitem{ATLASdil}
ATLAS Collaboration, Measurement of the top-quark mass in $\ttbar \rightarrow$dilepton events with the ATLAS experiment using the template method in 13 TeV $pp$ collision data, ATLAS-CONF-2022-058, \url{https://cds.cern.ch/record/2826701}

\bibitem{REVIEW}
Andr\'{e} Hoang, What Is the Top Quark Mass?, Ann. Rev. Nucl. Part. Sci. 70 (2020) 225-255

\end{thebibliography}
\end{document}